\documentstyle[12pt,epsf,amsfonts]{article}
\newcommand{\nn}{\nonumber}

\newcommand{\be}{\begin{equation}}
\newcommand{\ee}{\end{equation}}
\newcommand{\bea}{\begin{eqnarray}}
\newcommand{\eea}{\end{eqnarray}}
\newcommand{\Cech}{\v{C}ech\ }
\newcommand{\cU}{{\cal U}}

\newcommand{\CC}{{\mathbb C}}
\newcommand{\RR}{{\mathbb R}}
\newcommand{\ZZ}{{\mathsf Z}}
\newcommand{\Ab}{{\mathsf A}}
\newcommand{\Zn}{{{\mathbb Z}_n}}
\newcommand{\Zt}{{{\mathbb Z}_2}}
\newcommand{\bZ}{{\mathbb Z}}

\newcommand{\spinc}{{\rm Spin^c}}
\newcommand{\Tr}{{\rm Tr}}
\newcommand{\Hol}{{\rm Hol}}

\newcommand{\ra}{\rightarrow}

\newcommand{\dS}{{\partial\Sigma}}
\newcommand{\tA}{{\tilde A}}
\newcommand{\tw}{{\tilde w}}
\newcommand{\Pfaff}{{\rm Pfaff}}
\newcommand{\ii}{{\sqrt{-1}}}
\newcommand{\tB}{{\tilde B}}
\newcommand{\Adj}{{\rm\bf Adj}}
\newcommand{\End}{{\rm End}}
\newcommand{\Mat}{{\rm Mat}}
\newcommand{\Cont}{{\rm Cont}}

\title{\vspace{-1in}\parbox{\linewidth}{\small\hfill
\shortstack{IASSNS-HEP-99/82}}
\vspace{0.6in}\\
\bf D-branes in a topologically nontrivial $B$-field}
  
\author{Anton Kapustin\thanks{email: kapustin@ias.edu}\\
{\small \it School of Natural Sciences, Institute for Advanced Study}\\
{\small \it Olden Lane, Princeton, NJ 08540}}

\begin{document}
\begin{titlepage}
\renewcommand{\thepage}{ }
\renewcommand{\today}{ }

\thispagestyle{empty}
\maketitle 

\begin{sloppypar} 
\begin{abstract} 
We study global worldsheet anomalies for open strings ending on several
coincident D-branes in the presence of a $B$-field. We show that
cancellation of anomalies is made possible by a correlation between
the t'Hooft magnetic flux on the D-branes and the topological class of the 
$B$-field. One application of our results is a proper understanding of
the geometric nature of the gauge field living on D-branes: rather than
being a connection on a vector bundle, it is a connection on a module
over a certain noncommutative algebra. Our argument works for a general
closed string background. We also
explain why in the presence of a topologically nontrivial $B$-field
whose curvature is pure torsion D-branes represent classes in a twisted
K-theory, as conjectured by E.~Witten.

\end{abstract} 
\end{sloppypar}
\end{titlepage}

\section{Introduction and summary}
The goal of this paper is to understand the conditions for cancellation of global
worldsheet anomalies for open strings ending on D-branes. In the case when all
D-branes are noncoincident, this has been achieved in~\cite{FW}. It turns out
that interesting new phenomena arise when one considers coincident D-branes
with nonabelian gauge fields. In particular, noncommutative geometry
makes an appearance. Recently, it has been argued by many authors that 
the gauge field on D-branes is better understood as a connection on a module over some
noncommutative algebra rather than as a connection on a vector bundle.
This has been demonstrated for D-branes in flat space in the presence
of a constant $B$-field~\cite{NG}. Our setup is much more general 
(arbitrary metric and $B$-field), so it is nice to see that the same interpretation of 
the gauge field arises again. On the other hand, we will be concerned only
with the topological properties of the gauge field, leaving its dynamics aside.

\subsection{Review of twisted K-theory}

The main motivation for this study was the interpretation of
D-brane charges in terms of K-theory~\cite{MM,W}, so we begin by reviewing it.
It has been suspected for some time that D-brane charges are most naturally
understood as classes in K-theory rather than as integer cohomology classes~\cite{MM}.
The basic reason for this is that D-branes carry vector bundles. 
In Ref.~\cite{W} E. Witten showed that the relation between D-branes and K-theory
can be understood very simply using some ideas of A. Sen~\cite{sen}. 
For IIB string theory one considers a configuration of equal number of 
D9 and anti-D9-branes carrying vector
bundles $E$ and $F$. The pair $(E,F)$ defines a class in K-theory. It is 
easy to see that creation of virtual brane-antibrane pairs from vacuum does not change
the K-theory class. Thus the stable state into which the brane-antibrane
system settles after tachyons condense can be labeled by this class. It is
also very plausible that if pairs $(E,F)$ and $(E',F')$ have the same
K-theory class, then they lead to the same stable state, as one can be
converted to the other by a virtual process. Conversely, Witten 
explained how, given a D-brane wrapped on a submanifold, one can construct 
a K-theory class it represents.

In the presence of the Neveu-Schwarz B-field these arguments should be modified.
One way to see this is to note~\cite{W,FW} that a single D-brane can be wrapped
on a submanifold $Q$ if and only if the normal bundle $N$ of $Q$ satisfies:
\be\label{cond}
W_3(N)=[H]_Q.
\ee
Here $W_3(N)\in H^3(Q,\bZ)$ is the image of the second Stiefel-Whitney class
$w_2(N)$ under the Bockstein homomorphism, $[H]$ is the integer cohomology 
class whose de Rham representative is $H=dB$,
and $[H]_Q$ is its restriction to $Q$. 
Eq.~(\ref{cond}) is the condition of cancellation of global
anomalies for open strings ending on $Q$~\cite{FW}. It refines the usual statement
that the restriction of $H$ to the worldvolume of a D-brane must be trivial in
de Rham cohomology. When $[H]_Q=0$ the condition
(\ref{cond}) says that $N$ has $\spinc$ structure, and enables one to 
construct a class in the K-theory of space-time $X$ corresponding to a 
D-brane wrapped on $Q$~\cite{W}. When $[H]_Q$ is 
nonzero, the submanifold $Q$ does not represent a K-theory class. 

In Ref.~\cite{W} it has been proposed that when $[H]$ belongs to a torsion
subgroup of $H^3(M,\bZ)$, D-brane charges take values in a certain ``twisted''
version of K-theory $K_H(M)$. (In mathematical literature it is sometimes called
K-theory with local coefficients.) To motivate the definition of $K_H(M)$, let us
recall that the space of global sections $\Gamma(M,E)$ of a (finite-dimensional) vector 
bundle $E$ is a (finitely generated) projective module over the algebra of 
continuous $\CC$-valued functions on $M$ $\Cont_M(\CC)$. (A projective module is a 
module which is a direct summand of a free module. $\Gamma(M,E)$ is projective for any 
$E$ because any $E$ is a direct summand of a trivial vector bundle.)
Since $M$ is completely determined by the algebra $\Cont_M(\CC)$, it should be possible
to define $K(M)$ directly in terms of $\Cont_M(\CC)$. Such a definition is indeed possible:
for any algebra $\Ab$ one can define its K-group $K(\Ab)$ as the Grothendieck group
of finitely generated projective modules over $\Ab$. By a theorem of Serre-Swan~\cite{Se,Sw}
the category of finitely generated projective modules over $\Cont_M(\CC)$ is equivalent to 
the category of finite-dimensional vector bundles over $M$, hence 
$K(\Cont_M(\CC))=K(M)$.

Now it is plausible that when the B-field is present, we must replace $\Cont_M(\CC)$
with some other (noncommutative) algebra $\Ab$ and consider $K(\Ab)$. 
Morally speaking, this follows from the lore that when the B-field is switched 
on, $M$ becomes a noncommutative space in the sense of Connes~\cite{CDS,NG}.

A natural class of algebras is provided by so-called Azumaya algebras~\cite{Gro}. 
An Azumaya algebra of rank $m$ over $M$ is a locally trivial algebra bundle over 
$M$ whose fiber is isomorphic to the algebra of $m\times m$ matrices 
$\Mat_m(\CC)$.
A trivial example of an Azumaya algebra over $M$ is $\Mat_m(\CC)\otimes \Cont_M(\CC)$, 
i.e. the
algebra of $\Mat_m(\CC)$-valued functions on $M$. A slightly less trivial 
example is the algebra ${\rm End}(E)$ of all endomorphisms of a vector bundle $E$ over $M$.
Two Azumaya algebras $\Ab$ and $\Ab'$ over $M$ are said to be equivalent if there exist two
vector bundles $E$ and $E'$ over $M$ such that $\Ab\otimes {\rm End}(E)$ is isomorphic to 
$\Ab'\otimes {\rm End}(E')$. In particular, an Azumaya algebra of the form 
${\rm End}(E)$ is equivalent in this sense to $\Cont_M(\CC)$ for any $E$. 
This definition of equivalence
is useful because the K-group of an Azumaya algebra depends only on its equivalence
class~\cite{DK}. Moreover, by a theorem of Serre~\cite{Gro} equivalence 
classes of Azumaya algebras over $M$ are classified by the torsion subgroup
of $H^3(M,\bZ)$.\footnote{This subgroup is called the topological Brauer group
of $M$ and is denoted $Br(M)$.} We will denote by $\delta(\Ab)$ the class in $H^3(M,\bZ)$
corresponding to $\Ab$. To summarize, for any torsion 
class $[H]\in H^3(M,\bZ)$ there is a unique equivalence class of Azumaya algebras 
and the corresponding K-group $K_H(M)$. It is natural to conjecture~\cite{W} that 
$K_H(M)$ classifies D-brane charges when $[H]$ is a torsion class.

It possible to give a more down-to-earth description of Azumaya algebras and their modules.
Let us pick an open cover $\{ \cU_i\}$ such that all $\cU_i$ and their multiple overlaps
are contractible. An element of an Azumaya algebra $\Ab$ of rank $m$ is a section of a vector bundle
with fiber $\Mat_m(\CC)$, i.e. it is a collection of functions $R_i: \cU_i\ra \Mat_m(\CC)$
such that on double overlaps $\cU_i\cap \cU_j=\cU_{ij}$ we have
\be\label{zero}
R_i=G_{ij} R_j G_{ij}^{-1}.
\ee
Here $G_{ij}$ are $GL(m,\CC)$-valued functions over $\cU_{ij}$. Consistency of (\ref{zero})
on triple overlaps $\cU_{ijk}$ requires
\be\label{delta}
G_{ij} G_{jk} G_{ki}=\zeta_{ijk},
\ee
where $\zeta_{ijk}$ are $\CC^*$-valued functions. It is easy to see that $\zeta_{ijk}$ define
a \Cech cocycle, and therefore an element $\zeta\in H^2(M,\Cont_M(\CC^*))$. Here $\CC^*$ is the
group of nonzero complex numbers. $\zeta$ is a measure of
nontriviality of the Azumaya algebra $\Ab$. By a well-known isomorphism $H^2(M,\Cont_M(\CC^*))
\simeq H^3(M,\bZ)$
we can reinterpret $\zeta$ as a class in $H^3(M,\bZ)$; this is the class
$\delta(\Ab)$ mentioned above.

A module over an Azumaya algebra can be described as follows. Suppose we are given a 
set of functions $h_{ij}: \cU_i\ra GL(n,\CC)$
satisfying on triple overlaps
\be\label{twistzero}
h_{ij} h_{jk} h_{ki}=\zeta_{ijk},
\ee
where the cocycle $\zeta$ is the same as in (\ref{delta}). Consider a vector bundle
with fiber $\CC^{nm}$ and transition functions 
\be\label{transmodule}
h_{ij}^{-1\, T}\otimes G_{ij}.
\ee
First of all, such a bundle makes sense since the transition functions satisfy the usual 
``untwisted'' gluing condition on triple overlaps. Second, its fiber 
\be\label{module}
\CC^{nm}\simeq \CC^m\oplus\cdots\oplus\CC^m
\ee
is naturally a module over $\Mat_m(\CC)$, the fiber of $\Ab$. Third, it is easy to check that
the obvious ``fiberwise'' action of $\Ab$ makes it into a module over $\Ab$. It can be shown
that this module is projective if $M$ is a nice enough space (compact, for example).
Moreover, all projective modules of $\Ab$ arise in this way.\footnote{Let us give
a sketch of a proof. First, a projective module over $\Ab$ is also a projective 
module over $\Cont_M(\CC)$ and therefore corresponds to a vector bundle $E$ over $M$. Second, it
can be shown that the fiber of $E$ must be a module over $\Mat_m(\CC)$. Third, any module
over $\Mat_m(\CC)$ has the form (\ref{module}) for some $n$~\cite{rings}. Fourth, the 
action of $\Ab$ on any given fiber $E_x$ of $E$ generates a subalgebra $\Ab_x$ of 
$\End (E_x)\simeq \Mat_{nm}(\CC)$ isomorphic to $\Mat_m(\CC)$. Fifth, let the transition functions for 
$E$ be $g_{ij}$. It is easy to see that in order for the fiberwise
action of $\Ab$ to be consistent on double overlaps, $G_{ij}^{-1} g_{ij}$ must be in the
centralizer $C(\Ab_x)$ of $\Ab_x$. Since both $\Ab_x$ and $\End(E_x)$ are central simple algebras,
by Corollary 3.16 of \cite{rings} $\End(E_x)\simeq \Ab_x\otimes C(\Ab_x)$, and it follows that
there exists a trivialization of $E$ such that $g_{ij}$ have the form (\ref{transmodule}).} 

\subsection{Azumaya algebras and D-branes}

Now we are going to explain the relation between global worldsheet anomalies
and Azumaya modules. The basic idea is rather simple.
The starting point for the Sen-Witten construction of stable D-branes in IIB string 
theory is a configuration containing $n$ D9-branes and $n$ anti-D9-branes.
When $[H]=0$, the D-branes carry a principal bundle with structure group $U(n)$.
We will show that when $[H]$ is nonzero, the D-branes carry a principal $SU(n)/\Zn$ 
bundle which cannot be lifted to a $U(n)$ bundle. The obstruction to lifting an $SU(n)/\Zn$
bundle to a $U(n)$ bundle is a certain class in $H^3(M,\bZ)$ closely related to the
t'Hooft magnetic flux. We will prove that cancellation of global worldsheet anomalies
requires that this class be equal to $[H]$. 

An $SU(n)/\Zn$ bundle without $U(n)$ structure can described in terms of transition
functions as follows. Take an associated vector bundle in the adjoint representation of 
$SU(n)/\Zn$. Its section is given by a collection of functions $f_i: \cU_i\ra u(n)$,
where $u(n)$ is the Lie algebra of $SU(n)/\Zn$, i.e. the space of all hermitian $n\times n$
matrices. On double overlaps they satisfy
\be
f_i=h_{ij} f_j h_{ij}^{-1},
\ee
where $h_{ij}$ are $U(n)$-valued matrices.
Consistency on triple overlaps requires 
\be\label{twist}
h_{ij} h_{jk} h_{ki}=\zeta_{ijk},
\ee
where $\zeta_{ijk}$ are $U(1)$-valued functions.
The cohomology class of $\zeta$ in $H^2(M,\Cont_M(U(1)))\simeq H^3(M,\bZ)$ is the obstruction 
to lifting the $SU(n)/\Zn$ structure to a $U(n)$ structure.  

Comparing (\ref{twist}) and (\ref{twistzero}) we discover that an $SU(n)/\Zn$ bundle
without $U(n)$ structure corresponds naturally to a module over an Azumaya algebra
$\Ab$ with $\delta(\Ab)=\zeta$. Since we have both D9-branes and anti-D9-branes, we get a pair
of modules over the same algebra $\Ab$ which define a class in $K_H(M)$. It is also
easy to see that creation and annihilation of branes does not change this class.

Furthermore, the gauge field in the fundamental
representation of $U(n)$ (whose holonomy enters the open string path integral,
see below) cannot be a connection on a rank $n$ vector bundle
over $M$, since there is no such bundle around. Instead, the gauge field
is a connection on the Azumaya module defined above.

Thus cancellation of global worldsheet anomalies for open strings ending on D9-branes is
the reason why D-brane charges take values in $K_H(M)$. More generally, one can 
analyze these anomalies for open strings ending on D-branes of arbitrary 
dimension. This yields consistency conditions for wrapping multiple D-branes on
a submanifold $Q\subset M$. To formulate these consistency conditions we need to be more 
precise about the relation between the t'Hooft magnetic flux and the obstruction to having 
$U(n)$ structure.
The t'Hooft magnetic flux is a cohomology class $y\in H^2(Q,\Zn)$ which measures the obstruction 
to lifting an $SU(n)/\Zn$ structure to an $SU(n)$ structure. Consider a short exact sequence 
of groups
\be\label{exact}
0\ra \bZ \ra \bZ \ra \Zn \ra 0,
\ee
where the second arrow is multiplication by $n$ and the third arrow is reduction modulo $n$.
It leads to a long exact sequence in cohomology 
\be
\ldots H^2(Q,\Zn)\ra H^3(Q,\bZ) \ra H^3(Q,\bZ)\ldots
\ee
The homomorphism $H^2(Q,\Zn)\ra H^3(Q,\bZ)$ is a Bockstein homomorphism and will be 
denoted $\beta'$. As explained in more detail below, the obstruction to having a $U(n)$ structure
is precisely $\beta'(y)$. We will show that for bosonic strings cancellation of global 
anomalies requires 
\be\label{bosoncond}
\beta'(y)=[H]_Q.
\ee
For superstrings we find that global anomalies cancel if
\be\label{super}
\beta'(y)+W_3(N)=[H]_Q.
\ee
This generalizes (\ref{cond}). We will also explain the analogue of these conditions for Type I 
D-branes.

\subsection{The mechanism of anomaly cancellation}

Showing that global worldsheet anomalies
cancel if (\ref{bosoncond})  or (\ref{super}) are satisfied is unfortunately rather
tedious. (For an analogous argument in the
abelian case see~\cite{FW}, Section 6. The argument for Type I D9-branes
was sketched in \cite{SS}.) This is because we tried to keep
the discussion elementary yet reasonably rigorous. Let us briefly explain what is involved in the
argument.
For simplicity, consider the path integral for an open bosonic string in the presence of D9-branes:
\be\label{integral}
\int {\cal D} \xi\ \exp\left[\ii S_{NG}[\xi]+\ii \int_\Sigma \xi_* B\right]\ \Tr\ 
\Hol_{\partial\Sigma} (A).
\ee
Here $\xi$ is the map from the worldsheet $\Sigma$ to the target space $M$, 
$S_{NG}[\xi]$ is the Nambu-Goto action, $A$ is the $U(n)$ gauge field on the D9-branes, 
$\Hol_\gamma(A)$ is the holonomy of $A$ along the loop $\gamma$, and $\Tr$ is the
trace taken in the fundamental representation of $U(n)$.
It will be sufficient to consider the case when $\Sigma$ is a disc.
The Nambu-Goto action is a well-defined function on the space of maps 
$Map(\Sigma,M)$ from $\Sigma$ to $M$, but the other two factors are more problematic.
Indeed, while $H=dB$ is a well-defined 3-form, $B$ itself may have ``Dirac string 
singularities.'' Similarly, since the $SU(n)/\Zn$ bundle cannot be lifted to a 
$U(n)$ bundle, the trace of holonomy of
the gauge connection in the fundamental representation of $U(n)$ is not well-defined.
This can be traced back to the ``twisted'' gluing condition (\ref{twist}).

In fact, both of these factors make perfect sense if we interpret them as sections
of certain line bundles over $Map(\Sigma,M)$ rather than as mere functions.
Recall that the trace of holonomy of a connection on an ordinary 
vector bundle is a function on a free 
loop space $LM$ of $M$. We can define the trace of holonomy for our ``connection'' 
as well, but it turns out that it takes values in a certain nontrivial line bundle 
over $LM$. Similarly, the phase factor coming from the $B$-field is a section of
a well-defined line bundle over $Map(\Sigma,M)$. 

We are actually interested not in holonomy around arbitrary loops $S^1\ra M$, but
around those loops which extend to a map $\Sigma\ra M$ where $\Sigma$ is a disc.
Thus we are interested not in the above-mentioned line bundle over $LM$ but in its 
pull-back to $Map(\Sigma,M)$. We will show that the pull-back is trivial, and the phase 
factor coming from the B-field provides its trivialization. As a result, although 
neither the trace of holonomy nor the phase coming from the B-field are functions on 
$Map(\Sigma,M)$, their product is a well-defined function. This implies that the
open string path integral is well-defined.

\subsection{Outline of the paper}

The paper is organized as follows. In Section II we recall how to
define $B$-field in a topologically nontrivial situation. In Section III we describe
the topology of the gauge fields living on D-branes. In Section IV we explain how
one is supposed to understand the trace of holonomy of a connection in the fundamental
representation of $U(n)$ in the presence of t'Hooft magnetic flux. In Section V we
define the phase factor $\exp (-\ii \int \xi_*B)$ following \cite{Alv,Ga} and show that the open string path
integral for bosonic strings is well-defined if the condition (\ref{bosoncond}) is satisfied.
In Section VI we extend the discussion to superstrings and show that anomalies cancel
if (\ref{super}) is satisfied. As explained above, this implies that stable D-branes in Type IIB
string theory are classified by $K_H(X)$. In Section VII we discuss what happens when
$[H]$ is not a torsion class. 

A word about conventions. We denote by $\RR$ the field of real numbers,
and by $\ZZ$ the abelian subgroup of $\RR$ generated by $2\pi$. 
Of course, $\ZZ$ is canonically isomorphic to the group of integers $\mathbb Z$.
It is also a module over the ring $\mathbb Z$.
We will normalize $H=dB$ so that its periods are integer multiples of $2\pi$. 
Thus $[H]$ will be a class in $H^3(M,\ZZ)$ rather than in $H^3(M,{\mathbb Z})$.

\section{The definition of the B-field}

The precise definition of the B-field is most conveniently formulated in the 
language of \v{C}ech---de Rham cohomology~\cite{Alv,FW}.
Choose an open cover $\cU_i, i\in I,$ of $M$, such that all $\cU_i$, the double overlaps 
$\cU_{ij}=\cU_i\cap\cU_j$, the triple overlaps $\cU_{ijk}$, etc., are  
contractible (such a cover exists for any manifold.)
On each ${\cU}_i$ there is a well-defined real 2-form $B_i$.
On double overlaps $\cU_{ij}$ we have the relation
\be\label{one}
B_j-B_i=d\Lambda_{ij},
\ee
where $\Lambda_{ij}$ are real 1-forms. These 1-forms satisfy a 
consistency condition on triple overlaps $\cU_{ijk}$:
\be\label{two}
\Lambda_{ij}+\Lambda_{jk}+\Lambda_{ki}=-\ii d\log\alpha_{ijk},
\ee
where $\alpha_{ijk}$ are $U(1)$-valued 0-forms. In turn, $\alpha_{ijk}$ satisfy a 
further constraint on quadruple overlaps $\cU_{ijkl}$:
\be\label{three}
\alpha_{ijk}\alpha_{jkl}^{-1}\alpha_{ikl}\alpha_{ijl}^{-1}=1.
\ee
Equations (\ref{one}-\ref{three}) mean that $B$-field is 
a connection on a gerbe~\cite{Bry}, as defined in~\cite{Hi}. However, we will not use this 
terminology here.

Equation~(\ref{three}) implies
\be\label{five}
-\log\alpha_{ijk}+\log\alpha_{jkl}-\log\alpha_{ikl}+\log\alpha_{ijl}=\ii m_{ijkl},
\ee
where $m_{ijkl}\in\ZZ$. $m_{ijkl}$ form a \Cech cochain with values in $\ZZ$.
Moreover, it is easy to see from (\ref{five}) that $m$ is a \Cech cocycle and therefore 
represents a class in $H^3(M,\ZZ)$. This class will be called $[H]$, because its de 
Rham representative is $H=dB$.

Note that the condition (\ref{three}) means that the functions $\alpha_{ijk}$ form a 2-cocycle
with values in $\Cont_M(U(1))$, the sheaf of $U(1)$-valued continuous functions on $M$.
One can construct the corresponding class in $H^2(M,\Cont_M(U(1))$ directly
from $[H]$ by the following standard argument. Consider an exponential exact 
sequence of sheaves
\be\label{sheaves}
0\ra \ZZ\ra \Cont_M(\RR)\ra \Cont_M(U(1))\ra 0.
\ee
It induces a long exact sequence in cohomology which reads in part
\be\label{fine}
\ldots H^2(M,\Cont_M(\RR))\ra H^2(M,\Cont_M(U(1)))\ra H^3(M,\ZZ)\ra 
H^3(M,\Cont_M(\RR))\ldots.
\ee
But the sheaf $\Cont_M(\RR)$ is fine, so $H^p(M,\Cont_M(\RR))=0$ for all $p>0$. It
follows that $H^2(M,\Cont_M(U(1)))$ is isomorphic to $H^3(M,\ZZ)$. The cohomology class 
of $\alpha$ is mapped to $[H]$ under this isomorphism.

We are interested in the situation when we are given a submanifold $Q\subset M$
and the restriction of $H$ to $Q$ is trivial in de Rham cohomology, i.e. $[H]_Q$ is 
pure torsion. Until Section~\ref{cga}
we are going to work exclusively with objects on $Q$, so in order not to clutter
notation we will omit the subscript $Q$ until further notice. Hopefully
this will not cause confusion. 

An open cover $\{\cU_i\}$ of $M$ induces an open cover of $Q$ indexed by the same set $I$.
By abuse of notation, we will continue to refer to the elements of the cover as $\cU_i$.
If $H$ is trivial in de Rham cohomology, there exists a 2-form $\tB$ such that $H=dB$.
Then $B_i=\tB+\mu_i$, where $\mu_i$ is a real 1-form on $\cU_i$. 
{}From (\ref{one}) it follows that on double overlaps we
have \be\label{six}
\mu_j-\mu_i-\Lambda_{ij}=d\rho_{ij},
\ee
for some real 0-forms $\rho_{ij}$. {}From (\ref{two}) and (\ref{six}) it follows that
on triple overlaps we have
\be\label{seven}
-\ii (\rho_{ij}+\rho_{jk}+\rho_{ki})=\log \alpha_{ijk}-\log\zeta_{ijk},
\ee
where $\zeta_{ijk}$ are complex numbers with unit modulus. The numbers
$\zeta_{ijk}$ form a \Cech cochain with values in $U(1)$. 
In fact (\ref{five}) implies that $\zeta$ is a \Cech cocycle and so defines
a class in $H^2(Q,U(1))$. The origin of this class can be understood more abstractly
as follows. Consider an exponential exact sequence of groups
\be\label{first}
0\ra \ZZ \ra \RR \ra U(1) \ra 0.
\ee
It leads to a long exact sequence in the cohomology of $Q$ which reads in part
\be
\ldots H^2(Q,U(1))\ra H^3(Q,\ZZ) \ra H^3(Q,\RR)\ldots
\ee
The first arrow here is a Bockstein homomorphism which we will call $\beta$.
Since the cohomology class of $H$ in $H^3(Q,\RR)$ is trivial, exactness
implies that there is an element $\zeta\in H^2(Q,U(1))$ such that $\beta(\zeta)=[H]$.
$\zeta$ is precisely the class we have constructed above using \Cech cocycles.

\section{The topology of the gauge bundle on D-branes.}

It is usually said that $A$ (the gauge field on D-branes) is a
connection on a $U(n)$ vector bundle over $Q$, but transforms nontrivially under the 
gauge transformation $B\ra B+d\Lambda$, to wit $A\ra A-\Lambda$. We need a more precise 
statement which would tell us how 1-forms $A_i$ on various patches are glued together. 
We regard $A$ as a collection of 1-forms $\tA_i, i\in I.$ 
We postulate that on double overlaps they satisfy
\be\label{transf1}
\tA_i=g_{ij} \tA_j g_{ij}^{-1}+\ii g_{ij}d g_{ij}^{-1}-\Lambda_{ij}.
\ee
Here $\Lambda_{ij}$ are the same 1-forms as in (\ref{one}), and $g_{ij}$ are
$U(n)$-valued functions. The transformation law (\ref{transf1}) is not
the correct transformation law for a connection on a vector bundle. 
It is not even immediately clear how to define the holonomy of such a
``connection'' $A$ around a loop. A natural definition exists only when the
restriction of $H$ to $Q$ is trivial in de Rham cohomology.
It will turn out that the trace of holonomy of $A$ takes values in a line bundle over
the loop space $LQ$ rather than in complex numbers. 

Apart from the last term in (\ref{transf1}), the transformation
law for $\tA_i$ is that of a connection on a vector bundle. To be precise, for a
vector bundle the gluing functions $g_{ij}$ must satisfy
\be\label{glueord}
g_{ij} g_{jk} g_{ki}=1.
\ee 
We will impose instead 
\be\label{gluetwistp}
g_{ij} g_{jk} g_{ki}=\alpha_{ijk},
\ee
where $\alpha_{ijk}$ were defined in (\ref{two}).
The reason for this peculiar definition is that it will make the open string path
integral well defined.

In order to understand in what sense $A$ is a ``connection'', we need to 
get rid of the last term in (\ref{transf1}). 
Recall that when $B$ is flat, $\Lambda_{ij}$ satisfy (\ref{six}). We define new 1-forms
$A_i$ by $$A_i=\tA_i-\mu_i.$$ Their transformation law follows from (\ref{transf1})
and (\ref{six}):
\be\label{transf2}
A_i=h_{ij} A_j h_{ij}^{-1}+\ii h_{ij} dh_{ij}^{-1},
\ee
where 
\be
h_{ij}=g_{ij} e^{\ii \rho_{ij}}.
\ee
The transformation law for $A_i$ looks like that of an ordinary connection on
a vector bundle. However, the gluing functions $h_{ij}$ do not satisfy the usual
condition (\ref{glueord}) on triple overlaps. Instead one can show, using 
(\ref{gluetwistp}),(\ref{seven}), and (\ref{five}) that they satisfy
\be\label{twisttwo}
h_{ij} h_{ij} h_{ki}= \zeta_{ijk}.
\ee
Thus $h_{ij}$ do not define a $U(n)$ vector bundle over $Q$. Instead, as explained in 
Section I, they define a module $\Gamma$ over an Azumaya algebra $\Ab$ with 
$\delta(\Ab)=\beta(\zeta)=[H]$.

Although the transition functions $h_{ij}$ taken in the fundamental representation
of $U(n)$ do not define a vector bundle, they do define a vector bundle when taken
in the adjoint representation of $U(n)$. This is because $h_{ij}$ fail to glue
properly only modulo the elements of the center of $U(n)$, which is immaterial 
once one passed to the adjoint representation. Thus we obtain a rank $n^2$ bundle
over $Q$ which we call $\Adj$. It is easy to see that the 1-forms $A_i$ lead to a 
well-defined connection on $\Adj$. This enables one to define the holonomy of $A$ in 
the adjoint representation in the usual manner.

The bundle $\Adj$ has $U(n)/U(1)=SU(n)/\Zn$ as its
structure group. One
can define a class $y$ in $H^2(Q,\Zn)$ which measures the obstruction to lifting
it to an $SU(n)$ bundle. This class is called the t'Hooft magnetic flux in the
physics literature. Let us construct a \Cech cocycle representing $y$. Set
\be\label{eight}
q_{ij}=\left(\det h_{ij}\right)^{1/n},
\ee
and define $SU(n)$-valued transition functions
\be\label{nine}
{\tilde h}_{ij}=q_{ij}^{-1}h_{ij}.
\ee
In general, they do not glue properly over triple overlaps; instead we have
\be\label{ten}
{\tilde h}_{ij}{\tilde h}_{jk}{\tilde h}_{ki}=y_{ijk}.
\ee
Taking the determinant of both sides of (\ref{ten}) we see that 
$y_{ijk}$ is a 2-cochain with values in $\Zn$. Furthermore, it follows
directly from the definition of $y$ that it is a 2-cocycle. This cocycle is
the \Cech representative of the t'Hooft magnetic flux.

For our purposes it is more useful to know if $\Adj$ can be lifted to a $U(n)$ bundle.
This may be possible even if $y\neq 0$. Indeed, suppose $y$ is nonzero, but
is a reduction modulo $n$ of an integer cohomology class $z$. This means that there
exists a $\ZZ$-valued \Cech cocycle $z_{ijk}$ such that
\be
y_{ijk}=\exp\left(\frac{\ii z_{ijk}}{n}\right).
\ee
A standard argument shows
there exists a line bundle ${\cal L}$ over $Q$ whose first Chern class is 
$z$.\footnote{First one uses the exact sequence (\ref{fine}) and the fact that
$\Cont_Q(\RR)$ is a fine sheaf to show that $H^1(Q,\Cont_Q(U(1)))\simeq H^2(Q,\ZZ)$.
Let $t$ be the preimage of $z\in H^2(Q,\ZZ)$ under this isomorphism.
The \Cech cocycle $t_{ij}$ representing $t$ yields the transition functions
for ${\cal L}$.}
Let its transition functions be $t_{ij}$, i.e.
\be
\log t_{ij}+\log t_{jk}+\log t_{ki}=-\ii z_{ijk}.
\ee
Then the $U(n)$-valued functions $u_{ij}={\tilde h}_{ij} t_{ij}^{1/n}$ satisfy
\be
u_{ij}u_{jk}u_{ki}=1.
\ee
This means that $u_{ij}$ define a vector bundle with structure group $U(n)$.
It follows that a sufficient condition for being able to lift $\Adj$ to
a $U(n)$ bundle is
\be\label{mod}
\exists z\in H^2(Q,\ZZ)\quad y=z\ {\rm mod}\ 2\pi n.
\ee
It is easy to see that this condition is also necessary.

There is a convenient way to rewrite this criterion. Consider a short exact sequence
of groups
\be\label{exactthree}
0\ra \ZZ\ra \ZZ\ra \Zn\ra 0,
\ee
where the second arrow is multiplication by $n$. This exact sequence leads to
a long exact sequence in cohomology
\be\label{exactfour}
\ldots H^2(Q,\ZZ)\ra H^2(Q,\Zn)\ra H^3(Q,\ZZ)\ra H^3(Q,\ZZ)\ldots.
\ee
The second arrow here is a Bockstein homomorphism which we call $\beta'$.
Exactness implies that $y$ is in the kernel of $\beta'$ if and only if it satisfies
(\ref{mod}). Thus the necessary and sufficient condition for being able to
lift $\Adj$ to a $U(n)$ bundle is
\be
\beta'(y)=0.
\ee

We now claim that $\beta'(y)=[H]$. Indeed, it easily follows from our definitions
that  
\be\label{zyrelation}
\zeta_{ijk}=y_{ijk} q_{ij} q_{jk} q_{ki}.
\ee
Taking the logarithm of
both sides of (\ref{zyrelation}) yields
\be
\log\zeta_{ijk}=\log y_{ijk}+\log q_{ij}+\log q_{jk}+\log q_{ki}+\ii p_{ijk},
\ee 
where $p_{ijk}$ are $\ZZ$-valued numbers. Next we apply the \Cech coboundary operator to
both sides and get 
\be\label{Bock}
-\log\zeta_{ijk}+\log\zeta_{jkl}-\log\zeta_{ikl}+\log\zeta_{ijl}=
-\log y_{ijk}+\log y_{jkl}-\log y_{ikl}+\log y_{ijl}+\ldots,
\ee
where the dots denote a \Cech 3-coboundary with values in $\ZZ$.
The left-hand side is a \Cech cocycle representing $\beta(\zeta)\equiv [H]$, while
the right-hand side is a \Cech cocycle representing $\beta'(y)$ plus
a coboundary. Thus we get
\be
\beta'(y)=[H].
\ee

An important consequence of this relation is $n[H]=0$, i.e. $[H]$ is an 
$n$-torsion class. Indeed, the third arrow in (\ref{exactfour}) is multiplication
by $n$, so exactness implies $n\beta'\equiv 0$, hence $n[H]=0$.

Let us summarize what we have learned so far. The postulated transformation law for the
``connection 1-forms'' $\tA_i$ (\ref{transf1}) leads to a well-defined vector bundle 
$\Adj$ of rank $n^2$ with structure group $SU(n)/\Zn$. The 1-forms $A_i=\tA_i-\mu_i$ 
taken in the adjoint representation of $U(n)$ define a good connection on $\Adj$. In general 
$\Adj$ carries a nontrivial t'Hooft magnetic flux $y\in H^2(Q,\Zn)$ which prevents lifting
it to an $SU(n)$ bundle. Even if $y\neq 0$, it may still be possible to lift $\Adj$
to a $U(n)$ bundle; the obstruction to doing this is $\beta'(y)\in H^3(Q,\ZZ)$.
We showed that the Bockstein of $y$ is precisely $[H]$. This means that  
$\Adj$ can be lifted to a $U(n)$ bundle if and only if $[H]$ is trivial.
(Caution: remember that all objects here are restricted to $Q$, i.e.
$[H]$ is really $[H]_Q$.)

\section{The definition of the trace of holonomy}

In the previous section we remarked that the 1-forms $A_i$ enable one to compute the holonomy
and its trace in the adjoint representation of $U(n)$. What we really need, however,
is the trace of holonomy in the fundamental
representation. 

We have seen that when $\beta(\zeta)\equiv [H]$ is nonzero, it is impossible to lift
the $SU(n)/\Zn$ bundle $\Adj$ to a $U(n)$ bundle. In such a situation the 1-forms $A_i$ cannot be 
interpreted as connection 1-forms for a $U(n)$ vector bundle. Nevertheless, they do have a 
geometric meaning: they define a connection on an Azumaya module $\Gamma$. Roughly speaking, 
a connection on an Azumaya module $\Gamma$ is a connection on the underlying vector bundle which 
is compatible with the action of $\Ab$ on $\Gamma$. A more precise definition goes as follows. 
Suppose we are given a connection on an Azumaya algebra $\Ab$. This means that we are given a 
covariant derivative $\nabla^\Ab$ on the underlying vector bundle such that for any two elements 
$R_1,R_2\in \Ab$ and any vector field $X$ we have
\be
\nabla^\Ab_X(R_1R_2)=\nabla^\Ab_X(R_1)R_2+R_1\nabla^\Ab_X(R_2).
\ee
Then a connection on a module $\Gamma$ over $\Ab$ is a covariant derivative $\nabla^\Gamma$ on the 
underlying vector bundle such that for any $s\in \Gamma, R\in \Ab$ and any vector field $X$ we have
\be\label{Leibniz}
\nabla^\Gamma_X(Rs)=\nabla^\Ab_X(R)\, s+R\,\nabla^\Gamma_X(s).
\ee

In terms of an open cover $\{\cU_i\}$ and transition functions~(\ref{zero}), a connection on $\Ab$ 
is given by
\be
\nabla^\Ab_\mu(R)_i=\partial_\mu R_i-\ii\,\left[b_{i\mu}, R_i\right],
\ee
where $b_i=b_{i\mu}dx^\mu$ are $u(m)$-valued 1-forms which satisfy on double overlaps 
\be
b_i=G_{ij} b_j G_{ij}^{-1}+\ii\, G_{ij} dG_{ij}^{-1}.
\ee
Then a connection on a module $\Gamma$ is given by
\be
\nabla^\Gamma_\mu (s)_i=\partial_\mu s_i - \ii\, (1\otimes b_{i\mu}-A_{i\mu}^T\otimes 1)s_i,
\ee
where $A_i$ are the $u(n)$-valued 1-forms defined in the previous section.
The property (\ref{Leibniz}) is easily verified.

A crucial difference between a connection on the module $\Gamma$ and a connection
on an ordinary vector bundle arises when one tries to define the holonomy of a connection 
along a loop $\gamma\in LQ$. {}From (\ref{twisttwo}) one readily sees that any natural 
definition of holonomy will produce an object which is well-defined only modulo 
multiplication by a complex number with unit modulus. For this reason we anticipate that 
the trace of holonomy of $A$ will actually take values in a line bundle over $LQ$.

To define the trace of holonomy of $A$ we will work locally on $LQ$.
We have already picked an open cover $\cU_i, i\in I,$ of $Q$. An open cover of $LQ$ 
will consist of sets $V_p, p\in {\cal P}$. ${\cal P}$ is the set of pairs $(t,f)$,
where $t$ is a triangulation of $S^1$ and $f$ is a map from the simplices
of $t$ to $I$. By definition $V_{(t,f)}\subset LQ$ is the set of loops $\gamma: S^1\ra Q$ such that
for any simplex $\sigma\in t$ $\gamma(\sigma)\subset \cU_{f(\sigma)}$. Now pick any
$p=(t,f)\in {\cal P}$. Let $t$ consist of $r$ simplices
$\sigma_1,\ldots,\sigma_r$. Since $S^1$ is oriented, the simplices are also oriented. 
Let $v_\ell,\ell=1,\ldots,r$ be the vertices of $t$ numbered
so that the boundary of $\sigma_\ell$ is $v_{\ell+1}-v_\ell$. We will use
a shorthand $f_\ell=f(\sigma_\ell)$. 

Suppose we are given a $\gamma \in V_p$. We denote by $A_\ell$ the pull-back of the
1-form $A_{f_\ell}$ to $\sigma_\ell$. We denote by $h_{\ell,\ell-1}$
the matrix $h_{f_\ell f_{\ell-1}}(\gamma(v_\ell))$.
We now define a function ${\cal F}_p: V_p\ra \CC$ by the following formula:
\be
{\cal F}_p(\gamma)=\Tr\left[\prod_{\ell=n}^{1} 
\Hol_{\sigma_\ell}(A_\ell) h_{\ell,\ell-1}\right],
\ee
where $\Tr$ is the trace in the fundamental representation.

Suppose the intersection $V_p\cap V_{p'}=V_{pp'}$ is nonempty. We want to
compare ${\cal F}_p$ and ${\cal F}_{p'}$ on the double overlap $V_{pp'}.$
Let $\gamma\in V_{pp'}$; then a short computation shows that
\be\label{Ftrans}
{\cal F}_p(\gamma)={\cal H}_{pp'}{\cal F}_{p'}(\gamma),
\ee
where ${\cal H}_{pp'}$ is a $U(1)$-valued 1-cochain on $LQ$. In order to
write it down we need to introduce some more notation. 
Let $\sigma'_1,\ldots, \sigma'_{r'}$ be the simplices
of $t'$. We define a nondecreasing sequence of integers $i_0,\ldots,i_r$
by $$v_1\in \sigma'_{i_1},\quad v_2\in \sigma'_{i_2},\quad \ldots,\quad
v_r\in \sigma'_{i_r}.$$ The cochain ${\cal H}$ is given by
\bea\label{cocy1}
&&\left(\prod_{j_r=i_r}^{j_r=i_{r-1}+1} \zeta_{f'_{j_r}f_r f'_{j_r-1}}\right)
\zeta_{f'_{i_{r-1}}f_r f_{r-1}}  
\left(\prod_{j_{r-1}=i_{r-1}}^{j_{r-1}=i_{r-2}+1}
\zeta_{f'_{j_{r-1}}f_{r-1}f'_{j_{r-1}-1}}\right)\ldots \nn\\ 
&&\left(\prod_{j_1=i_2}^{j_r=i_1+1} \zeta_{f'_{j_1}f_1 f'_{j_1-1}}\right)
\zeta_{f'_{i_1}f_1 f_r}.
\eea
{}From (\ref{Ftrans}) it follows that on triple overlaps $V_{pp'p''}$ the transition functions 
${\cal H}_{pp'p''}$ satisfy
\be
{\cal H}_{pp'}{\cal H}_{p'p''}{\cal H}_{p''p}=1.
\ee 
In other words, ${\cal H}$ is a 1-cocycle over $LQ$ with values in
$U(1)$. 

The cocycle ${\cal H}$ defines a line bundle ${\cal L}_A$ over $LQ$. 
The first Chern class of this bundle $c_1({\cal L}_A)\in H^2(LQ,\ZZ)$ is the image 
of ${\cal H}$ under the Bockstein homomorphism $\beta'':H^1(LQ,U(1))\ra H^2(LQ,\ZZ)$.
Since the transition functions ${\cal H}_{pp'}$ are constant, the line bundle 
${\cal L}_A$ is flat and its first Chern class is pure torsion.

We define the trace of holonomy of $A$ in the fundamental representation to be
the global section of the line bundle ${\cal L}_A$ which is given locally by
${\cal F}_p$. Thus the trace of holonomy of $A$ is a section of a flat 
line bundle over $LQ$. 

\section{Cancellation of global anomalies for bosonic strings}\label{cga}
In this section we establish that the path integral for an open bosonic string ending
on D-branes is well-defined if the transition functions $h_{ij}$ satisfy the twisted gluing
condition (\ref{twisttwo}).

Recall that $\xi$ maps $\dS$ to a submanifold $Q\subset M$ on which the D-branes
are wrapped. {}From now on we will reinstate subscripts $Q$ where necessary. We assume
that $[H]_Q$ is a torsion class.

Consider the space $Map(\Sigma,M)$ of maps from the worldsheet $\Sigma$ to $M$
such that $\dS$ is mapped to a submanifold $Q\subset M$. We are dealing with oriented
strings, so $\Sigma$ and $\dS$ are oriented. It will be sufficient to consider the
case when $\dS$ has a single component, i.e. $\Sigma$ is a Riemann surface with a single
hole. If we choose a diffeomorphism $S^1\ra \dS$, there is a natural projection
$\pi:Map(\Sigma,M)\ra LQ$. 
In the previous section we showed that the trace of holonomy of $A$ is a section of 
a certain line bundle ${\cal L}_A$ over $LQ$. This bundle is flat, but in general nontrivial.
For string theory applications we need to consider the pull-back of ${\cal L}_A$ to 
$Map(\Sigma,M)$ by $\pi$. It turns out that the pull-back is 
trivial, and its trivialization is given by $\exp (-\ii\int \xi_*B).$ 
To prove this, we should first accurately define $\exp (-\ii\int \xi_*B).$
This has been done for worldsheets without boundaries
in \cite{Alv} and extended to worldsheets with boundaries in \cite{Ga}.
Below we describe the construction of \cite{Alv,Ga} as it applies to our particular
situation.
A version of this construction which does not depend on the choice of open cover
is explained in Chapter 6 of \cite{Bry}.

We warn the reader that the accurate definition of $\exp (-\ii\int \xi_*B)$ is rather
lengthy, so on first reading he or she may want to jump to the next section. The upshot
is that $\exp (-\ii\int \xi_*B)$ is a trivialization of the line bundle $\pi_*({\cal L}_A)$. 
Moreover,
our definition of $\exp (-\ii\int \xi_*B)$ is such that the product 
of the last two factors in (\ref{integral}) is a well-defined function on $Map(\Sigma,M)$.

\subsection{The case of a closed worldsheet}

We will start by defining $\exp (- \ii \int \xi_*B)$ when $\Sigma$ is a closed
oriented surface. In this case we want $\exp (-\ii \int \xi_*B)$ to be a
well-defined function on $Map(\Sigma,M)$. Moreover, we expect that this function
is invariant with respect to gauge transformations $B\ra B+d\lambda$, where $\lambda$
is a globally defined 1-form.

Recall that we have a good cover of $M$ indexed by a set $I$. Let ${\cal S}$ be the set 
of pairs $(\tau,\phi)$ where $\tau$ is the triangulation of
$\Sigma$ and $\phi$ is a map from the simplices of $\tau$ to $I$. For any $s\in {\cal S}$ we 
define a set $W_s\subset Map(\Sigma,M)$ consisting of all the maps $\xi:\Sigma\ra M$
such that for any simplex $s\in \tau$ $\xi(s)\subset \cU_{\phi(s)}$. The sets $W_s, s\in {\cal S}$
form an open cover of $Map(\Sigma,M)$. 

We will work locally on $Map(\Sigma,M)$.
We want to define a collection of functions $\Phi_s: W_s\ra U(1)$ indexed by elements
of ${\cal S}$ and show that they agree on double overlaps $W_{ss'}=W_s\cap W_{s'}$. 
To do this we have to 
introduce some more notation. Let us fix a triangulation $\tau$ of $\Sigma$. Let $\{s_a\}, 
a\in {\cal A}$, be the set of all simplices of $\tau$, let $\{e_b\}, b\in {\cal B}$, be the
set of all edges, and let $\{v_c\}, c\in {\cal C}$ be the set of all vertices.
Any edge belongs to exactly two simplices. A vertex can belong to two or more simplices.
We will say that $v$ is $n$-valent if it belongs to exactly $n$ simplices.

An orientation of $\Sigma$ induces an orientation on all the simplices of $\tau$. 
An edge $e_{aa'}=s_a\cap s_{a'}$ can be oriented if we specify an ordering of $a,a'$.

Let us fix a map $\phi:{\cal A}\ra I$.
Let $\xi\in W_{(\tau,\phi)}$. Then we can use $\xi$ to pull back a 2-form $B_\phi(a)$ on 
$\cU_{\phi(a)}$ to a 2-form $B_a$ on the simplex $s_a$. Similarly, we can pull back
a 1-form $\Lambda_{\phi(a)\phi(a')}$ to a 1-form $\Lambda_b$ on the edge $e_b$, where $e_b$ 
is shared by the simplices $s_a$ and $s_{a'}$. Finally, suppose we have a vertex
$v$ shared by simplices $s_{a_1},\ldots,s_{a_n}$. For any three indices
$a,a',a''\in \{a_1,\ldots,a_n\}$ we will define a complex number $\alpha_{aa'a''}$ as
$\xi_*\alpha_{\phi(a)\phi(a')\phi(a'')}(v)$. 

Given $s=(\tau,\phi)$ and $\xi\in W_s$ we now define a $U(1)$-valued function on $W_s$:
\be
\tilde\Phi_s(\xi)=\exp \left[ -\ii\sum_{a\in {\cal A}}\int_{s_a} B_a-\ii\sum_{b\in{\cal B}} 
\int_{e_b} \Lambda_b\right]. 
\ee
The functions $\tilde\Phi_s$ are not quite what we need because they do not agree on
double overlaps. However, this can be fixed by introducing a correction factor 
$C_v$ for each vertex of the triangulation which depends on the 2-cocycle $\alpha$ from
(\ref{two}). The correction factor is defined as follows.

1.  If a vertex $v$ is divalent, $C_v=1$. 

2. Suppose $v$ is trivalent. Since $\Sigma$ is oriented, the simplices sharing $v$
have a natural cyclic order $s_a,s_{a'},s_{a''}$. We set
$C_v=\alpha_{aa'a''}.$

3. Suppose the vertex $v$ belongs to $n$ simplices $s_{a_1},\ldots,s_{a_n}$, $n>3$. 
We pick an arbitrary index $a\in \{a_1,\ldots,a_n\}$ and set
\be
C_v=\alpha_{a_1a_2a}\alpha_{a_2a_3a}
\ldots\alpha_{a_na_1a}.
\ee
It is easy to check that $C_v$ does not depend on the choice of $a$ because of
$(\ref{three})$. 

A patient reader should be able to prove that the functions $\Phi_s$ thus defined
agree on double overlaps $W_{ss'}$. The proof uses the relations
(\ref{one})-(\ref{three}). We define $\exp (-\ii \int \xi_*B)$ to be the global
function which is given locally by $\Phi_s$. It is obvious that the global function
thus defined is invariant with respect to gauge transformations $B\ra B+d\lambda$.

\subsection{The case of a worldsheet with a boundary}

Let $\Sigma$ be a Riemann surface with a single hole such that the boundary $\dS$ is mapped
to $Q$. For any triangulation $\tau$ we divide the edges and vertices into internal and boundary ones.
The sets ${\cal B}$ and ${\cal B}'$ will label the internal and boundary edges, respectively,
while the sets ${\cal C}$ and ${\cal C}'$ will label the internal and boundary vertices, respectively.
The set ${\cal A}$ will label the simplices of $\tau$, as before.
A boundary edge belongs to exactly one simplex and has a natural orientation induced by 
that of $\dS$. A boundary vertex belongs to one or more simplices. Suppose
that a boundary vertex $v$ belongs to $n$ simplices $s_{a_1},\ldots,s_{a_n}$. Orientation of $\dS$
induces a natural order on $a_1,\ldots,a_n$ (not just cyclic order!)

We define functions $\Phi_s:W_s\ra U(1)$ as a product of ``internal'' and ``boundary'' contributions.
The internal contribution is the same as for a closed $\Sigma$. 
To define the boundary contribution we need to introduce some more notation. 
We use $\xi$ to pull back 1-forms 
$\mu_i$ on $\cU_i$ to 1-forms $\mu_b, b\in {\cal B}'$ on $e_b$. Similarly,
to every boundary vertex shared by the simplices $s_{a_1},\ldots,s_{a_n}$
we associate a set of numbers $\rho_{aa'}$ where $a,a'\in \{a_1,\ldots,a_n\}$ by pulling
back 0-forms $\rho_{ij}$ (see (\ref{six}). The boundary contribution is a product of
a factor associated with boundary edges and a factor associated with boundary vertices.
The first factor is
\be
\exp \left[\ii\sum_{b\in{\cal B}'}\int_{e_b} \mu_b\right].
\ee
The second factor is a product over all boundary vertices; a factor $C_v$ associated with an
individual vertex is defined by the following rules:

1. If the vertex $v$ is univalent, $C_v=1$.

2. If the vertex $v$ is divalent and belongs to an ordered pair of simplices $s_a$ and 
$s_{a'}$, then $C_v=\exp\left(\ii\rho_{aa'}\right).$

3. Suppose the vertex $v$ is $n$-valent, $n>2$. Let $s_{a_1},\ldots,s_{a_n}$ be the
ordered simplices sharing $v$. We set
\be
C_v=\alpha_{a_3 a_2 a_1}\alpha_{a_4 a_3 a_1}\ldots
\alpha_{a_n a_{n-1}a_1}\exp\left(\ii\rho_{a_1 a_n}\right).
\ee

The functions $\Phi_s:W_s\ra U(1)$ are now completely defined. It remains to compute the
transition functions on double overlaps. A very patient reader should be able
to prove that
\be\label{cocytwo}
\Phi_s={\cal H}_{\omega(s)\omega(s')}\Phi_{s'},
\ee
where ${\cal H}$ is the 1-cocycle (\ref{cocy1}) and $\omega$ is a natural projection 
$\omega:{\cal S}\ra {\cal P}$ induced by the natural projection $\pi: Map(\Sigma,M)\ra LQ$. 
Recall that ${\cal H}$ defined a line bundle ${\cal L}_A$ over $LQ$.
It follows that the collection
of functions $\{\Phi_s\}, s\in {\cal S}$ define a section of the pull-back line bundle 
$\pi_*({\cal L}_A)$. 

We now define $\exp (- \ii \int \xi_*B)$ as a global section of $\pi_*({\cal L}_A)$
which is given locally by $\Phi_s$. Since $\Phi_s$ take values in $U(1)$, $\exp (\ii \int \xi_*B)$
is a trivialization of $\pi_*({\cal L}_A^{-1})$. {}From (\ref{cocytwo}) and (\ref{Ftrans})
it follows that
\be
\exp \left(\ii \int \xi_*B\right)\ \Tr\ \Hol_{\dS}(\xi_* A)
\ee 
is a well-defined function on $Map(\Sigma,M)$. Hence the open string path integral is well-defined.
Moreover, it is easy to see that this function is invariant with respect to gauge transformations
$B\ra B+d\lambda, A\ra A-\lambda$, where $\lambda$ is a globally defined 1-form.

The above argument can be trivially extended to the case when $\dS$ has several
components. The details are left as an exercise for the reader.

\section{Generalizations}

We are of course much more interested in superstrings rather than bosonic strings.
In superstring theory $M$ is a spin manifold, while $Q\subset M$ is not necessarily spin.
As explained in \cite{FW}, the superstring path integral contains an extra problematic
factor $\Pfaff(D)$, the pfaffian of the Dirac operator on the worldsheet. This object is
a section of a line bundle over the space
\be
X=Met(\Sigma)\times Map(\Sigma,M)
\ee
where $Met(\Sigma)$ is the space of metrics on $\Sigma$. We will call this line bundle
$\Pfaff$. $\Pfaff$ is flat, but in general nontrivial. 

One of the results of \cite{FW} is the computation of the Chern class of $\Pfaff$. 
Namely, the first Chern class is given by
$$c_1(\Pfaff)=\pi'_* {\cal T} (W_3(N)).$$ 
Here $\pi'$ is natural projection $\pi': X\ra LQ$, ${\cal T}$ is the 
transgression homomorphism~\footnote{Let us remind the definition of
the transgression homomorphism. Let $x$ be a class in $H^3(Q,\ZZ)$. Consider the evaluation
map $e: S^1\times LQ\ra Q$ given by $e:(t,\gamma)\mapsto \gamma(t)$. We can use $e$ to pull back 
$x$ to $S^1\times LQ$. The transgression of $x$ is a class in $H^2(LQ,\ZZ)$
by obtained by ``integrating $e_*(x)$ over $S^1$.'' More formally, given a homology class $a\in H_2(LQ,\ZZ)$
we define ${\cal T}(x)(a)=e_*(x)(w\times a)$, where $w\in H_1(S^1,\bZ)$ is the fundamental homology class
of $S^1$.} 
${\cal T}: H^3(Q,\ZZ)\ra H^2(LQ,\ZZ)$, and $W_3(N)$ is
the Bockstein of the second Stiefel-Whitney class $w_2(N)$ of the normal bundle of $Q$.
Since $W_3(N)$ is a 2-torsion class, $c_1(\Pfaff)$ is also a 2-torsion
class.

In Section III we have also explained that the trace of holonomy of a connection on
an Azumaya module takes values in a line bundle ${\cal L}_A$ over $LQ$. We have explicitly
computed its transition functions in terms of a 2-cocycle $\zeta$ on $Q$. Since the
transition functions ${\cal H}_{pp'}$ are constant, ${\cal L}_A$ is flat, but in general
nontrivial. A natural question is whether one can express the Chern class of ${\cal L}_A$ 
directly in terms of data on $Q$. The answer to this question can be inferred from the results
of \cite{Ga,Bry} where it is shown how to construct a line bundle ${\cal L}_H$ over $LQ$ given a 
class $[H] \in H^3(Q,\ZZ)$. The 1-cocycle (i.e. the transition functions) defining ${\cal L}_H$ 
is given in Eq.~(6-19) of \cite{Bry}. It is easy to see that when $[H]$ is a torsion class, 
this cocycle is cohomologous to our cocycle ${\cal H}$ and therefore ${\cal L}_H$ is 
isomorphic to ${\cal L}_A$. It is explained in~\cite{Ga,Bry} that the first Chern class
of ${\cal L}_H$ (and therefore ${\cal L}_A$) is the transgression of $[H]$.  

Consider now the product of $\Pfaff(D)$ and the trace of holonomy of $A$ in the fundamental
representation. The first factor is a section of the line bundle $\Pfaff$ over $X$.
The second factor is a section of a line bundle $\pi'_*({\cal L}_A)$ over $X$.
The Chern class of $\pi'_*({\cal L}_A)$ is $\pi'_*{\cal T}(\beta(\zeta))$, where $\zeta$ is
the 2-cocycle entering the twisted gluing condition (\ref{twisttwo}). In Section IV
we have shown that if $\beta(\zeta)\in H^3(Q,\ZZ)$ is a restriction of a class in
$H^3(M,\ZZ)$, then $\pi'_*{\cal L}_A$ is trivial, and therefore $\pi'_*{\cal T}(\beta(\zeta))=0$.
In this section we do not assume that $\beta(\zeta)\in H^3(Q,\ZZ)$ is a restriction of 
a class in $H^3(M,\ZZ)$, hence $\pi'_*{\cal T}(\beta(\zeta))$ need not vanish.
The product of $\Pfaff(D)$ and the trace of holonomy of $A$ takes values in a line bundle 
$\Pfaff\otimes \pi'_*({\cal L}_A)$ with first Chern class
\be
\pi'_*{\cal T}\left(W_3(N)+\beta(\zeta)\right).
\ee 
Cancellation of global worldsheet anomalies requires this class to vanish. {}From Section IV we know
that if
\be\label{supercond}
W_3(N)+\beta(\zeta)=[H]_Q
\ee
for some class $[H]=H^3(M,\ZZ)$, then $\Pfaff\otimes \pi'_*({\cal L}_A)$ is trivial and its 
trivialization is given by $\exp (- \ii \int \xi_*B)$. Thus (\ref{supercond}) is a sufficient
for cancelling the global worldsheet anomalies. It is very plausible that that it is also
a necessary condition. From Section III we know that $\beta(\zeta)=\beta'(y)$, where $y\in H^2(Q,\Zn)$
is the t'Hooft magnetic flux, therefore Eq.~({\ref{supercond}) is equivalent to Eq.~(\ref{super}).

In the special case of D9-branes the condition (\ref{super}) reduces to $\beta'(y)=[H]$. 
This means that if $[H]\neq 0$,
the bundle on D9-branes is an $SU(n)/\Zn$ bundle which cannot be lifted to a $U(n)$ bundle,
and the obstruction to the lift is precisely $[H]$. As explained in the introduction, this
implies that a system of equal number of D9-branes and anti-D9-branes defines a class
in $K_H(M)$. Arguments analogous to those in~\cite{W} then suggest that stable D-branes
in IIB string theory are classified by $K_H(M)$.

It is possible to extend the discussion to Type I strings. Type I strings are unoriented, so
the definition of $\exp (- \ii \int \xi_*B)$ given in Section IV must be modified. 
In addition, for Type I strings $H=dB$ vanishes and $[H]$ is always a 2-torsion class. We will
not go into the details of the anomaly cancellation mechanism for Type I strings, but it is
easy to guess what the analogue of (\ref{supercond}) should be. Consider first the case
when only D9-branes are present. They carry a $Spin(32)/\Zt$ bundle which in general
cannot be lifted to a $Spin(32)$ bundle. The obstruction to doing this is 
the ``generalized second Stiefel-Whitney class'' $\tw_2\in H^2(M,\Zt)$~\cite{six}.
However, in order to define the open string path integral, we need the trace of holonomy
in the vector representation of $Spin(32)$. We expect that this object is a section
of a line bundle over $LQ$ whose first Chern class is the transgression of $\beta'(\tw_2)$.
We thus expect that the path integral is well-defined if 
\be
\beta'(\tw_2)=[H],
\ee
This relation was first noticed in \cite{SS}.

Consider now $n$ Type I D5-branes wrapped on a submanifold $Q\subset M$. It is usually said that
Type I D5-branes carry a principal bundle with structure group $Sp(n)$. We expect that this 
statement will be modified when $[H]_Q\neq 0$ and the bundle will be an $Sp(n)/\Zt$ bundle 
without $Sp(n)$ structure. The obstruction to having $Sp(n)$ structure is measured by a class
$y\in H^2(Q,\Zt)$, and we expect that cancellation of anomalies requires
$W_3(N)+\beta'(y)=[H]_Q$. 

\section{Discussion}

In this paper we have shown that when the restriction of $[H]$ to the
D9-brane worldvolume is nonzero, cancellation of worldsheet anomalies forces the 
D9-brane bundle to be an $SU(n)/\Zn$ bundle without $U(n)$ structure. As a consequence of this 
the geometric meaning of the gauge field changes: it becomes a connection on a module over 
an Azumaya algebra. This fact is
responsible for the appearance of twisted K-groups $K_H$ in string theory.

More generally, we showed that cancellation of global worldsheet anomalies imposes a 
correlation between the t'Hooft magnetic flux on D-branes and $[H]_Q$. This
correlation is summarized in (\ref{bosoncond}) and (\ref{super}).

A natural question is whether it is possible to extend the present discussion
to the case when $[H]$ is not a torsion class. Witten has argued~\cite{W} that the
answer must be negative. We give two more arguments supporting this conclusion.

First, it must be clear from our discussion that when $[H]$ is not a torsion class,
it is impossible to wrap any finite number of D9-branes on $M$. This means that
there is no good starting point for the Sen-Witten construction of D-branes.

Second, even assuming that one can somehow make sense of an infinite number of
D9-branes (and anti-D9-branes), there seems to be no good candidate for a K-theory
for general $[H]$. Indeed, for any $[H]$ one would like to find an algebra bundle 
over $M$ such that its K-theory classifies the D-brane charges. The structure group
of this algebra bundle should be some suitable version of $SU(n)/\Zn$ as $n\ra \infty$.
A natural class of algebra bundles is suggested by a theorem of
Dixmier and Douady~\cite{DD,Bry} which states that algebra bundles whose fiber is 
the algebra of compact (or Hilbert-Schmidt, or trace
class) operators on a separable infinite-dimensional Hilbert space $V$ are in one-to-one
correspondence with elements of $H^3(M,\ZZ)$. The corresponding ``gauge group''
is $PU(V)$, the projective unitary group of $V$.
Unfortunately, the corresponding K-theories are not the right ones.
For example, the K-group corresponding to the trivial class in $H^3(M,\ZZ)$ would be
the Grothendieck group of Hilbert bundles on $M$ with structure group $U(V)$, the unitary
group of $V$. But this K-group is very different from the expected answer ${\tilde K}(M)$: it 
is trivial for any manifold $M$ because all Hilbert bundles over $M$ are isomorphic to a 
trivial one. This happens because the ``gauge group''
$U(V)$ is contractible~\cite{Kui}. More generally, one can consider bundles
or sheaves of von Neumann algebras; however none of them seems to
lead to an acceptable K-theory, because their automorphism groups have the wrong
homotopy type. 
In order to get the correct K-theory, one has to work
with much smaller ``gauge groups.'' For example, one could try a subgroup of
the unitary group $U^c(V)\subset U(V)$ whose elements have the form $1+K$, where $K$ is compact 
(or Hilbert-Schmidt, or trace-class.) One can show that the K-group of Hilbert bundles
with structure group $U^c(V)$ is the same as ${\tilde K}(M)$ ~\cite{Sv,Palais}. Unfortunately, 
unlike $U(n)$, $U^c(V)$ does not have a center, so this K-theory does not have a twisted 
version.
\vskip 12pt
\noindent {\bf Acknowledgements: }I would like to thank Mikhail Khovanov for useful discussions. 
This work was supported by a DOE grant
DE-FG02-90ER40542.


\begin{thebibliography}{99}
\bibitem{FW} D.S.~Freed and E.~Witten, ``Anomalies in string theory with D-branes,''
hep-th/9907189.

\bibitem{NG} M.R.~Douglas and C.~Hull, ``D-branes and the noncommutative torus,''
JHEP {\bf 02}, 008 (1998), hep-th/9711165; V.~Schomerus,
``D-branes and deformation quantization,'' JHEP {\bf 06}, 030 (1999),
hep-th/9903205; for a more extensive list of references see 
N.~Seiberg and E.~Witten, ``String theory and noncommutative geometry,''
hep-th/9908142.

\bibitem{MM}R.~Minasian and G.~Moore,
``K-theory and Ramond-Ramond charge,'' JHEP {\bf 11}, 002 (1997);
hep-th/9710230.

\bibitem{W} E.~Witten, ``D-branes and K theory,'' JHEP {\bf 12}, 019 (1998); 
hep-th/9810188.

\bibitem{sen} A.~Sen,
``SO(32) spinors of type I and other solitons on brane-antibrane pair,''
JHEP {\bf 09}, 023 (1998); hep-th/9808141.

\bibitem{Se} J.P.~Serre, ``Modules projectifs et espaces fibr\'{e}s \`{a} fibre
vectorielle,'' Exp. {\bf 23}, S\'{e}minaire Dubreil-Pisot, Paris, 1958.

\bibitem{Sw} R.G.~Swan, ``Vector bundles and projective modules,'' Trans. Amer. Math. 
Soc. {\bf 105} (1962) 264-277.

\bibitem{CDS} A.~Connes, M.R.~Douglas and A.~Schwarz,
``Noncommutative geometry and matrix theory: Compactification on tori,''
JHEP {\bf 02}, 003 (1998); hep-th/9711162.

\bibitem{Gro} A.~Grothendieck, ``Le Groupe de Brauer I,'' S\'{e}minaire Bourbaki,
Exp. No. {\bf 290} (1964/1965) 01-21.

\bibitem{DK} P.~Donovan and M.~Karoubi, ``Graded Brauer groups and K-theory with
local coefficients,'' IHES Pub. {\bf 38} (1970) 5.

\bibitem{rings} B.~Farb and R.K.~Dennis, ``Noncommutative algebra,''
Graduate Texts in Mathematics {\bf 144}, Springer-Verlag, New York, 1993.

\bibitem{SS}A.~Sen and S.~Sethi, ``The Mirror transform of type I vacua in 
six-dimensions,'' Nucl. Phys. {\bf B499}, 45 (1997); hep-th/9703157. 

\bibitem{Alv} O.~Alvarez, ``Topological Quantization And Cohomology,''
Commun. Math. Phys. {\bf 100}, 279 (1985).

\bibitem{Ga} K.~Gawedzki,
``Topological Actions In Two-Dimensional Quantum Field Theories,''
in {\it Nonperturbative Quantum Field Theories}, eds. G.~t'Hooft, A.~Jaffe,
G.~Mack, P.K.~Mitter, and R. Stora, NATO Series {\bf 185}, Plenum Press, 1988,
p. 101-141.

\bibitem{Bry} J.-L.~Brylinski, ``Loop spaces, Characteristic Classes, and Geometric
Quantization,'' Progress in Mathematics {\bf 107}, Birkh\"{a}user, Boston, 1993.

\bibitem{Hi} N.~Hitchin, ``Lectures on special Lagrangian submanifolds,''
math.dg/9907034.

\bibitem{six} M.~Berkooz, R.G.~Leigh, J.~Polchinski, J.H.~Schwarz, N.~Seiberg and E.~Witten,
``Anomalies, dualities, and topology of D = 6 N=1 superstring vacua,''
Nucl. Phys. {\bf B475}, 115 (1996); hep-th/9605184.

\bibitem{DD} J.~Dixmier and A.~Douady, ``Champs continues d'espaces hilbertiens
et de $C^*$-alg\`{e}bres,'' Bull. Soc. Math. Fr. {\bf 91} (1963) 227-284.

\bibitem{Kui} N.H.~Kuiper, ``The homotopy type of the unitary group of Hilbert space,''
Topology {\bf 3} (1965) 19-30.

\bibitem{Sv} A.S.~Schwarz, ``On the homotopic topology of Banach spaces,'' 
Dokl. Akad. Nauk SSSR {\bf 154} (1964) 61-63.

\bibitem{Palais} R.~Palais, ``On the homotopy type of certain groups of operators,''
Topology {\bf 3} (1965) 271-279.




\end{thebibliography}
\end{document}